\documentclass{appolb}
\usepackage{array}
\usepackage{epsfig,epsf}
\usepackage{amsmath}
\usepackage{amsthm}
\usepackage{amsfonts}
\usepackage{amssymb}
\usepackage{dsfont}
\usepackage{multirow}
\usepackage{appendix}
\usepackage{slashed}
\usepackage{psfrag}
\usepackage{bbold}
\usepackage{graphicx}

\begin{document}

\begin{center}
{\large {\bf \sc{In the Pursuit of $X(5568)$ and its Charmed Partner}}} \\[2mm]
J.Y.~S\"ung\"u\footnote{Email: jyilmazkaya@kocaeli.edu.tr}, A.~T\"{u}rkan$^{2}$, E. Veli~Veliev$^{1}$\\[4mm]
$^{1}$ Department of Physics, Kocaeli University, 41380 Izmit, Turkey\\
$^{2}$ \"{O}zye\u{g}in University, Department of Natural and
Mathematical Sciences, \c{C}ekmek\"{o}y, Istanbul,Turkey\\
\end{center}

\begin{abstract}
The recent observation by the D$\O$ collaboration of the first
tetraquark candidate with four different quark flavors $(u, d, s$
and $ b)$ in the $B^0_s\pi^{\pm}$ channel having a narrow
structure, has still not been confirmed by other collaborations.
Further independent experiments are required either to confirm the
$X(5568)$ state or to set limits on its production. Though quantum
numbers are not exactly clear, the results existing in the
literature indicate that it is probably an axial-vector or scalar
state candidate. In this study, mass and pole residue of the
$X(5568)$ resonance assuming as a tightly bound diquark, with
spin-parity both $J^{P}=1^{+}$ or $J^{PC}=0^{++}$ are calculated
using two-point Thermal SVZ sum rules technique by including
condensates up to dimension six. Moreover, its partner in the
charm sector is also discussed. Investigations defining the
thermal properties of $X(5568)$ and its charmed partner may
provide valuable hints and information for the upcoming
experiments such as CMS, LHCb and PANDA.
\end{abstract}


\section{Introduction}

A new era began in the hadron spectroscopy in $2003$ when Belle
Collaboration announced the pioneering discovery of the enigmatic
resonance $X(3872)$~\cite{Choi:2003ue}. Since then there has been
an explosion in the discovery of exotic structures that cannot be
placed into the well-tested quark model of hadrons. This group of
particles are called XYZ states, to indicate their nature is
unclear, emerged from the Belle, BaBar, BESIII, LHCb, CDF, D$\O$
and other collaborations (for a review of these particles, see
Refs.~\cite{Godfrey:2008nc,Esposito:2014rxa,Liu:2013waa,Olsen:2014qna}).
The idea of the multiquark states was firstly put forward by Jaffe
in 1977 \cite{Jaffe:1976ig}. Especially after the observation of
$X(3872)$, this topic become very active research field in hadron
physics.

After thirteen years from this discovery, a unique structure
$X(5568)$ containing four different quark flavors such as
$[bd][\bar{s}\bar{u}], [bu][\bar{s}\bar{d}], [su][\bar{b}\bar{d}]$
or $[sd][\bar{b}\bar{u}]$ was reported by the D$\O$ Collaboration
in the decays $X(5568)\rightarrow B^0_s\pi^{\pm},~B^0_s\rightarrow
J/\psi\phi,~J/\psi\rightarrow \mu^+ \mu^-,~\phi\rightarrow
K^+K^-$. The exclusive features of the $X(5568)$ at the vicinity
of $D\overline{D}^*$ threshold, the tiny width and the large
isospin violation in production and decay, has opened up a new
window in hadron spectroscopy. Possible quantum numbers for this
state are $J^P = 0^+$, if the $B^0_s\pi^\pm$ is produced in an
S-wave or $J^P = 1^+$, if the decay proceeds via the chain
$X(5568)\rightarrow B^{*0}_s\pi^\pm,B^{*0}_s\rightarrow
B^0_s\gamma$ and the photon is not reconstructed. The measured
mass and width are $M_X = (5567.8 \pm
2.9\mathrm{(stat)^{+0.9}_{-1.9}(syst))~MeV}, \Gamma_X = (21.9 \pm
6.4\mathrm{(stat)^{+5.0}_{2.5}(syst))~MeV}$~\cite{D0:2016mwd},
respectively.

However the CDF and ATLAS Collaborations reported independently
negative search results for the $X(5568)$ state~
\cite{Aaltonen:2017voc,Aaboud:2018hgx}, while the D$\O$
Collaboration collected additional evidence by adding $B^0_s$
mesons reconstructed in semileptonic decays using the full run II
integrated luminosity of $10.4~fb^{-1}$ in $p\overline{p}$
collisions at a center of mass energy of $1.96~\mathrm{MeV}$ at
the Fermilab Tevatron Collider~\cite{Abazov:2017poh}. Further the
CMS Collaboration is accomplished a search for the $X(5568)$ state
by using $pp$ collision data collected at $\sqrt{s} =
8~\mathrm{TeV}$ and corresponding to an integrated luminosity of
$19.7~fb^{-1}$. With about $50000~B^0_s$ signal candidates, no
significant structure in the $B^0_s\pi^{\pm}$ invariant mass
spectrum is found around the mass reported by the D$\O$
Collaboration~\cite{Sirunyan:2017ofq}. Also, the LHCb
Collaboration did not confirm the existence of the
$X(5568)$~\cite{Aaij:2016iev}, which makes some theorists consider
the difficulty of explaining the $X(5568)$ as a genuine
resonance~\cite{Yang:2016sws,Albuquerque:2016isq,Lang:2016jpk}.

Although there exist different opinions on $X(5568)$,
re-observation of it in experiment ignites theorists enthusiasm of
surveying exotic tetraquark states. For instance in the framework
of QCD sum rule, Albuquerque and et al. investigated the $X(5568)$
state using the molecular interpolating currents
$BK,~B_s\pi,~B^*K,~B_s^*\pi$ and tetraquark currents with quantum
numbers $J^P = 0^+$ and $1^+$. Their numerical results did not
support the $X(5568)$ as a pure molecule or a tetraquark state.
However, they suggested it to be a mixture of $BK$ molecule and
scalar $[ds\bar{b}\bar{u}]$ tetraquark state with a mixing angle
$\mathrm{sin} 2\Theta\simeq 0.15$~\cite{Albuquerque:2016nlw}. Also
they conclude that $XZ$ states are good candidates for $1^+$ and
$0^+$ molecules or/and four-quark states while the predictions for
$1^-$ and $0^-$ states are about $1.5~\mathrm{GeV}$ above
$Y_{b,c}$ thresholds. To date, the resonance $X(5568)$ has
triggered lots of theoretical studies, most of which speculated it
to be a typical diquark-antidiquark state while the molecular
state assignment is not privileged.

The mass of $X(5568)$ is too far (nearly $200~\mathrm{MeV}$) below
from the $\overline{B}K$ threshold $(5774~\mathrm{MeV})$ to be
interpreted as a hadronic molecule of $\overline{B}K$.
Additionally, the interaction of $B^0_s\pi^\pm$ is very weak and
unable to form a bounded structure. The LHCb Collaboration scanned
the invariant mass of $B^0_s\pi^\pm$ and no significant signal for
a $B^0_s\pi^\pm$ resonance is seen at any value of mass and width
in the range considered~\cite{Aaij:2016iev}. Also the authors of
Ref.~\cite{Burns:2016gvy} deduced a lower limit for the masses of
a possible $[ds\bar{b}\bar{u}]$ tetraquark state: $6019$ MeV.
Completing Ref.~\cite{Burns:2016gvy}, Ref.~\cite{Guo:2016nhb}
presented an analysis based on general properties of QCD to
analyze the $X(5568)$ state. Notably, it was shown that the mass
of the $[ds\bar{b}\bar{u}]$ tetraquark state must be bigger than
the sum of the masses of the $B_s$ meson and the light
quark-antiquark resonance leading to an estimate of the lower
limit of $M_{bsud}\simeq 5.9~\mathrm{GeV}$. Moreover, in
Ref.~\cite{Agaev:2016mjb} and ~\cite{Agaev:2016lkl} mass of
$X_{b,c}$ are calculated both in axial-vector and scalar pictures,
respectively. In another work based on the same theory, i.e. QCD
sum rules, authors estimated the mass and decay constant of $X_b$
in scalar assumption computing up to the vacuum condensates of
dimension-10~\cite{Wang:2016mee} and in the charmed scalar sector
$D_{s0}(2317)$ was studied as the scalar tetraquark state,
too~\cite{Wang:2006uba}. The results obtained in this framework
were found to be nicely consistent with the experiments. Also in
Ref.~\cite{Ebert:2010af} mass of the $X_b$ ground state calculated
in the diquark-antidiquark picture in Relativistic Quark Model
(RQM) are higher than experimentally measured values as presented
in Tables~\ref{tab:Xbscalar} and Table~\ref{tab:Xbaxial} and in
the framework of Non-Relativistic Quark Model (NRQM) as well
\cite{Ghalenovi:2015mha}.

If the $X(5568)$ has a four-quark structure, its partner state
within the same multiplet must also exist. We assume that this
state bears the same quantum numbers as its counterpart, i.e.
$J^{P} = 1^{+}$ or $J^{PC} = 0^{++}$. We also accept that it has
the internal structure $X_c=[su][\bar{c}\bar{d}]$ in the
diquark-antidiquark model. Our aim is to determine the parameters
of the state $X_c$, i.e. to find its mass and pole residue. If
this partner state is not detected, one should put a big question
mark on the existence of the $X(5568)$ signal. According to
Ref.~\cite{Liu:2016ogz} charmed partner of the $X(5568)$ have more
strong decay channels than the bottom partners. Especially, the
experimental search for it are strongly called for in the
$D_s\pi$, $D^*_s\pi$, and isovector $\overline{D}\overline{K}$
channels. Due to explain its exotic decay modes, Liu et al. once
recommended a tetraquark structure for the $D_{sJ}(2632)$ signal
observed by the SELEX collaboration~\cite{Cooper:2004hj}. The mass
of this particle is very close to the $X_c$ meson. So this can be
the same particle with $X_c$. Unfortunately, $D_{sJ}(2632$) was
not confirmed by subsequent experiments.

Analyzing the thermal version~\cite{Bochkarev:1985ex} of this
ambiguous state $X(5568)$ using Shifman-Vainshtein-Zakharov Sum
Rule (SVZSR) model~\cite{Shifman} can give us a different point of
views. Hence, in this article we tentatively assume that $X(5568)$
and its charmed partner are exotic states and will focus on the
scenario of tetraquark state based on the SVZSR at finite
temperature using the deconfinement temperature $T_c=155
~\mathrm{MeV}$~\cite{Aoki:2006br,Andronic:2017pug,Steinbrecher:2018phh,Bazavov:2017dus}.
Our motivation for extension our computation to the high
temperatures is to interpret the heavy-ion collision experiments
more precisely. Moreover investigations of particles at finite
temperatures can give us information on understanding of the
nonperturbative dynamics of QCD, deconfinement and chiral phase
transition. We explore the variation of the mass and pole residue
values in terms of increasing temperature.

The article is arranged as follows. Section~\ref{sec:Theory} is
devoted to the description of the SVZSR approach at $T\neq0$. The
mass and pole residue sum rule expressions for the exotic
bottomonium and charmonium states are calculated by carrying out
the operator product expansion (OPE) up to condensates of
dimension-6. Then our numerical results for these quantities for
the relevant mesons are reported in Section~\ref{sec:NumAnal}.
Section~\ref{sec:Result} is reserved for our conclusions. Finally
the explicit forms of all spectral density expressions obtained in
the calculations are given in the Appendix.

\section{Thermal SVZ Sum Rule Formalism}\label{sec:Theory}

In this section we try to find the correlation function from both
the physical side (phenomenological side or hadronic side) and the
QCD side (OPE side or theoretical side). As stated in the SVZSR,
we can look at the quarks from both inside and also outside of the
hadrons, these two situation which is assumed as corresponding to
the same physical case can be calculated via two different
windows. Then equalizing the results coming from both sides, the
sum rules for the hadronic parameters are obtained.

Now assuming the $X(5568)$ state as a bound $[su][\bar{b}\bar{d}]$
tetraquark state and its charmed partner $X_c$ state as a
$[su][\bar{c}\bar{d}]$ tetraquark state, the mass and pole residue
sum rules of $X(5568)$ and $X_c$ resonances are obtained in hot
medium. In this study, Thermal SVZSR (TSVZSR) method is used
having applied to a wide range of hadronic observables from the
light to the heavy quark sector prosperously.

TSVZSR proposed by Bochkarev and Shaposnikov has been yielding a
brand-new research area
\cite{Bochkarev:1985ex,Hatsuda:1992bv,Alam:1999sc,Rapp:2009yu,Mallik:1997kj,Dominguez:2013fca,Veliev:2017fpa}.
The TSVZSR start with the two-point correlation function for the
scalar $\Pi(q,T)$ and axial-vector $\Pi_{\mu\nu}(q,T)$ assumption,
respectively:
\begin{equation}\label{CorrFuncScalar}
\Pi(q,T)=i\int d^{4}x~e^{iq\cdot x}\langle
\Psi|\mathcal{T}\Big\{\eta(x)\eta^{\dagger}(0)\Big\}|\Psi\rangle,
\end{equation}
\begin{equation}\label{CorrFuncAxial}
\Pi_{\mu \nu }(q,T)=i\int d^{4}x~e^{iq\cdot x}\langle
\Psi|\mathcal{T}\Big\{\eta_{\mu}(x)\eta_{\nu}^{\dagger}(0)\Big\}|\Psi\rangle,
\end{equation}
where $\Psi$ denotes the hot medium state, $\eta(x)$ and
$\eta_{\mu}(x)$ are the interpolating currents of the considered
particles and $\mathcal{T}$ represents the time ordered product
\cite{Shifman,Reinders,Colangelo:2000dp}. The thermal average of
any operator $\hat{O}$ in thermal equilibrium can be asserted by
the following expression:
\begin{equation}\label{operator}
\langle \hat{O}\rangle=\frac{Tr(e^{-\beta \mathcal{H}}\hat{O})}{Tr(e^{-\beta \mathcal{H}})}, \\
\end{equation}
where $\mathcal{H}$ is the QCD Hamiltonian, and being the $T$ is
the temperature of the heat bath, $\mathcal{\beta}=1/T$ is inverse
temperature.

Chosen currents $\eta(x)$ and $\eta_{\mu}(x)$ must contain all the
information of the related meson, like quantum numbers, quark
contents and so on. In the following, we will consider the
tetraquark states with quark contents $[su][\bar{b}\bar{d}]$ and
$[su][\bar{c}\bar{d}]$. In the diquark-antidiquark model currents
for the scalar and axial-vector states can be expressed as
\cite{Chen:2016mqt,Agaev:2016mjb,Agaev:2016lkl}:
\begin{eqnarray}\label{TetraCurrent}
\eta(x)&=&\epsilon_{ijk}\epsilon_{imn} \Big[s_{j}(x)C \gamma_{\mu}
u_{k}(x)\Big]\Big[\overline{Q}_{m}(x)\gamma_{\mu} C
\overline{d}_{n}(x)\Big],\nonumber \\
\eta_{\mu}(x)&=&s^T_j (x)C \gamma_5 u_k(x)\Big[~ \overline{Q}_j(x)
\gamma_{\mu} C \overline{d}^T_k(x)-\overline{Q}_k(x) \gamma_{\mu}
C \overline{d}^T_j(x)\Big],
\end{eqnarray}
respectively, where $Q=b$ or $c$ represent heavy quarks, $C$ is
the charge conjugation and $i,j,k,m,n$ are color indexes.

\subsection{\textbf{Physical Side}}

First we focus on the evaluation of the physical side of the
correlation function in order to determine the mass and pole
residue sum rules of $X(5568)$ and its charmed partner (hereafter
we will symbolize $X(5568)$ as $X_b$ and the charmed partner as
$X_c$). To derive mass and pole residue TSVZSR, we begin with the
correlation function with regard to the hadronic degrees of
freedom. Then we embed the complete set of intermediate physical
states possessing the same quantum numbers as the interpolating
current. Later, carrying out the integral over $x$ in Eqs.
(\ref{CorrFuncScalar}) and (\ref{CorrFuncAxial}), the following
expressions are obtained for the scalar and axial-vector
assumptions, respectively:
\begin{eqnarray}
\Pi^{\mathrm{Phys}}(q,T)&=&\frac{\langle
\Psi|\eta|X_{b(c)}(q)  \rangle \langle X_{b(c)}(q)  |\eta^{\dagger }|\Psi\rangle }{m_{X_{b(c)}}^{2}(T)-q^{2}}\nonumber \\
&+& higher~states,
\end{eqnarray}
\begin{eqnarray}
\Pi _{\mu \nu }^{\mathrm{Phys}}(q,T)&=&\frac{\langle
\Psi|\eta_{\mu }|X_{b(c)}(q)  \rangle \langle X_{b(c)}(q)
|\eta_{\nu
}^{\dagger }|\Psi\rangle}{m_{X_{b(c)}}^{2}(T)-q^{2}}\nonumber \\
&+& higher~states,
\end{eqnarray}
here $m_{X_{b(c)}}(T)$ is the temperature-dependent mass of
$X_{b(c)}$. Temperature dependent pole residues $f_{X_{b(c)}}(T)$
are defined with the following matrix elements:
\begin{equation}\label{ResidueScalar}
\langle \Psi|\eta|X_{b(c)}(q)  \rangle
=f_{X_{b(c)}}(T)~m_{X_{b(c)}}(T),
\end{equation}
\begin{equation}\label{ResidueAxial}
\langle \Psi|\eta_{\mu }|X_{b(c)}(q)  \rangle
=f_{X_{b(c)}}(T)~m_{X_{b(c)}}(T)~\varepsilon _{\mu},
\end{equation}
here $\varepsilon _{\mu}$ is the polarization vector of the
$X_{b(c)}$ state satisfying the following relation:
\begin{eqnarray}\label{polarizationvec}
\varepsilon_{\mu}\varepsilon_{\nu}^{*}=\frac{q_\mu
q_\nu}{m_{X_{b(c)}^2}(T)}-g_{\mu\nu}.
\end{eqnarray}
Then the correlation function depending on $m_{X_{b(c)}}(T)$ and
$f_{X_{b(c)}}(T)$ can be written in the below forms for the scalar
case
\begin{equation}\label{CorMScalar}
\Pi^{\mathrm{Phys}}(q,T)=\frac{m_{X_{b(c)}}^{2}(T)f_{X_{b(c)}}^{2}(T)}
{m_{X_{b(c)}}^{2}(T)-q^{2}}+\ldots
\end{equation}
and the axial-vector case:
\begin{eqnarray}\label{CorMAxial}
\Pi_{\mu\nu}^{\mathrm{Phys}}(q,T)&=&\frac{m_{X_{b(c)}}^{2}(T)f_{X_{b(c)}}^{2}(T)}
{m_{X_{b(c)}}^{2}(T)-q^{2}} \Big(\frac{q_{\mu }q_{\nu
}}{m_{X_{b(c)}}^{2}(T)}-g_{\mu \nu }\Big)+\ldots,
\end{eqnarray}
respectively. To obtain the TSVZSR we select a structure
consisting of $g_{\mu\nu}$ for the axial one from the
$\Pi_{\mu\nu}^{\mathrm{Phys}}(q,T)$, then using the coefficients
of this structure and applying the Borel transformation,
\begin{eqnarray}\label{Borel}
\mathcal{\hat B}_{(q^2)}[\Pi(q^2)]\equiv \lim\limits_{\substack{n\rightarrow \infty\\
}} \frac{(-q^2)^{n}}{(n-1)!}\Big(\frac{d^n}{{d}
q^{2n}}\Pi(q^2)\Big)_{q^2=n/M^2},
\end{eqnarray}
which improves the convergence of the OPE series and also enhances
the ground state contribution. So the physical side for the scalar
and axial-vector cases are acquired as;
\begin{eqnarray}\label{BorelPhys}
\mathcal{\hat{B}}_{(q^2)}[\Pi^{\mathrm{Phys}}(q,T)]=m_{X_{b(c)}}^{2}(T)f_{X_{b(c)}}^{2}(T)~e^{-m_{X_{b(c)}}^{2}(T)/M^{2}}.~
\end{eqnarray}\\
%
\subsection{\textbf{QCD Side}}

In this part, our purpose is to find the correlation function
belonging to the QCD side. $\Pi^{\mathrm{QCD}}(q,T)$
can be defined according to quark-gluon degrees of freedom. %
Similar to physical side, the correlation functions given in
Eqs.~(\ref{CorrFuncScalar}) and (\ref{CorrFuncAxial}) on the QCD
side are expanded in terms of Lorentz structures as well
\begin{eqnarray}\label{OPE_S}
\Pi^{\rm QCD}(q,T)&=&\Gamma_0(q,T)~\mathrm{I},\nonumber\\
\Pi^{\rm QCD}_{\mu\nu}(q,T)&=&\Gamma_{1}(q,T) g_{\mu\nu}+
\mbox{other structures},\quad
\end{eqnarray}
where $\Gamma_{0,1}(q,T)$ are the scalar functions in the Lorentz
structures that are selected in this work. In the rest frame of
the particle, related correlation functions are expressed with the
dispersion integral,
\begin{eqnarray}
\Pi^{\mathrm{QCD}}(q,T)=\int_{\mathcal{M}^{2}}^{s_{0}(T)}\frac{\rho
^{\mathrm{QCD}}(s,T)}{(s-q^{2})}ds+...,
\end{eqnarray}
where $\mathcal{M}=m_{s}+m_{u}+m_{b(c)}+m_{d}$ and the related
spectral density can be expressed as
\begin{eqnarray}\label{eq.QCDSidePxN}
\rho^{QCD}(s,T)=\frac{1}{\pi}Im[\Pi^{\mathrm{QCD}}(s,T)].
\end{eqnarray}
After briefly giving the general definitions, now we can start to
the computation for the OPE side placing the current expressions
in Eq.~(\ref{TetraCurrent}) into the QCD correlation function in
Eqs.~(\ref{CorrFuncScalar}) and (\ref{CorrFuncAxial}) and then
contracting the heavy and light quark fields, we have the
following expressions for the scalar assumption:
\begin{eqnarray}\label{eq:CorrFuncScalar}
\Pi^{\mathrm{QCD}}(q,T)&=&i\widetilde{\epsilon}\epsilon\int
d^{4}x~e^{iq\cdot x}\bigg[\mathrm{Tr}[\gamma _{\nu}\widetilde{S}
_{s}^{jj'}(x) \gamma _{\mu
}S_{u}^{kk'}(x)]\nonumber\\
&+&\mathrm{Tr}[\gamma _{\mu}\widetilde{S}_{d}^{n'n}(-x) \gamma
_{\nu }S_{b}^{m'm}(-x)] \bigg],
\end{eqnarray}
and the axial-vector one:
\begin{eqnarray}\label{eq:CorrFuncAxial}
\Pi_{\mu\nu}^{\mathrm{QCD}}(q,T)&=&i\widetilde{\epsilon}\epsilon\int
d^{4}x~e^{iq\cdot x}\bigg[\mathrm{Tr}[\gamma
_5\widetilde{S}_{s}^{jj'}(x) \gamma _5
S_{u}^{kk'}(x)]\mathrm{Tr}[\gamma_{\mu}\widetilde{S}_{d}^{j'k}(-x)\gamma_{\nu}
\widetilde{S}_{b}^{k'j}(-x)\nonumber\\
&-&\mathrm{Tr}[\gamma_5\widetilde{S}_{s}^{jj'}(x) \gamma_5
S_{u}^{kk'}(x)]\mathrm{Tr}[\gamma_{\mu}\widetilde{S}
_{d}^{k'k}(-x) \gamma _{\nu} S_{b}^{j'j}(-x)]\nonumber\\
&-&\mathrm{Tr}[\gamma_5\widetilde{S}_{s}^{jj'}(x) \gamma_5
S_{u}^{kk'}(x)]\mathrm{Tr}[\gamma_{\mu}\widetilde{S}_{d}^{j'j}(-x)
\gamma_{\nu}S_{b}^{k'k}(-x)]\nonumber\\
&+&\mathrm{Tr}[\gamma_5\widetilde{S}_{s}^{jj'}(x)\gamma_5
S_{u}^{kk'}(x)]\mathrm{Tr}[\gamma_{\mu}
\widetilde{S}_{d}^{k'k}(-x)\gamma_{\nu} S_{b}^{j'k}(-x)] \bigg],
\end{eqnarray}
where the notation $\widetilde{S}^{jj'}(x)=CS^{jj'T}(x)C$ is used
for brevity. Also, we have used the shorthand notations
$\epsilon=\epsilon_{ijk}\epsilon _{imn}$  and
$\tilde\epsilon=\epsilon_{i'j'k'}\epsilon _{i'm'n'}$ in Eqs.
(\ref{eq:CorrFuncScalar}) and (\ref{eq:CorrFuncAxial}). The quark
propagator in non-perturbative approach can be expressed with the
quark and gluon condensates~\cite{Reinders}. At finite temperature
additional operators arise since the breakdown of Lorentz
invariance by the choice of the thermal rest frame. The residual
$O(3)$ invariance naturally brings in additional operators to the
quark propagator in thermal case. The expected attitude of the
thermal averages of these new operators is opposite to those of
the Lorentz invariant old ones~\cite{Mallik:1997pq}.

General form of the heavy-quark propagator in this calculation in
the coordinate space can be expressed as in the below form:
\begin{eqnarray}\label{HeavyProp}
S_{Q}^{ij}(x)&=&i\int \frac{d^{4}k}{(2\pi )^{4}}e^{-ik\cdot x}\Bigg[ \frac{%
\delta _{ij}\Big( {\!\not\!{k}}+m_{Q}\Big)
}{k^{2}-m_{Q}^{2}}-\frac{gG_{ij}^{\alpha \beta }}{4}\frac{\sigma _{\alpha \beta }\Big( {%
\!\not\!{k}}+m_{Q}\Big) +\Big(
{\!\not\!{k}}+m_{Q}\Big)\sigma_{\alpha
\beta }}{(k^{2}-m_{Q}^{2})^{2}}\nonumber \\
&+&\frac{g^{2}}{12}G_{\alpha \beta }^{A}G_{A}^{\alpha \beta
}\delta_{ij}m_{Q}\frac{k^{2}+m_{Q}{\!\not\!{k}}}{(k^{2}-m_{Q}^{2})^{4}}+\ldots\Bigg],
\end{eqnarray}
where $G_{A}^{\alpha \beta}$ is the external gluon field, obeying
$G_{ij}^{\alpha \beta}=G_{A}^{\alpha \beta}t_{ij}^A$ with
\textit{A} is color indices from 1 to 8,
$t_{ij}^{A}=\lambda_{ij}^{A}/2$ and $\lambda_{ij}^{A}$ are the
Gell-Mann matrices. As for the thermal light-quark propagator
$S_{q}$ the following statement is employed:
\begin{eqnarray}\label{lightquarkpropagator}
S_{q}^{ij}(x) &=&i\frac{\slashed
x}{2\pi^{2}x^{4}}\delta_{ij}-\frac{
m_{q}}{4\pi^{2}x^{2}}\delta_{ij} -\frac{\langle \bar{q}q\rangle_T }{12}\delta_{ij}-\frac{x^{2}}{192}%
m_{0}^{2}\langle \bar{q}q\rangle_T \Big[1-i\frac{m_{q}}{6}\slashed x \Big]%
\delta _{ij}  \nonumber \\
&+&\frac{i}{3}\Big[\slashed x \Big(\frac{m_{q}}{16}\langle
\bar{q}q\rangle_T -\frac{1}{12}\langle u^{\mu} \Theta _{\mu \nu
}^{f} u^{\nu}\rangle \Big)  +\frac{1}{3}(u\cdot x)\slashed u
\langle u^{\mu}\Theta _{\mu \nu
}^{f} u^{\nu}\rangle %
\Big]\delta _{ij}  \nonumber \\
&-&\frac{ig_{s}G _{ij}^{\alpha\beta}}{32\pi ^{2}x^{2}}
\Big(\slashed x \sigma _{\mu \nu }+\sigma _{\mu \nu }\slashed
x\Big)-i\delta_{ij}\frac{x^2 \slashed x \langle \bar{q}q
\rangle^2_T}{7776} g_s^2 ,
\end{eqnarray}
where $m_{q}$ implies the light quark mass, $u_{\mu }$ is the
four-velocity of the heat bath, $\langle \bar{q}q\rangle_{T}$ is
the temperature-dependent light quark condensate being the $q=u,d$
or $s$ and $\Theta _{\mu \nu }^{f}$ is the fermionic part of the
energy momentum tensor. Also, for the gluon condensate with regard
to the gluonic part of the energy-momentum tensor $\Theta_{\lambda
\sigma }^{g}$, the consecutive relation is employed (see for
details Ref.~\cite{Mallik:1997pq}):
\begin{eqnarray}\label{TrGG}
&&\langle Tr^{c}G_{\alpha \beta }G_{\mu \nu }\rangle
=\frac{1}{24}(g_{\alpha \mu }g_{\beta \nu }-g_{\alpha \nu
}g_{\beta \mu })\langle G_{\lambda \sigma
}^{a}G^{a\lambda \sigma }\rangle   \nonumber  \label{TrGG} \\
&&+\frac{1}{6}\Big[g_{\alpha \mu }g_{\beta \nu }-g_{\alpha \nu
}g_{\beta \mu }-2(u_{\alpha }u_{\mu }g_{\beta \nu }-u_{\alpha
}u_{\nu }g_{\beta \mu }
\nonumber \\
&&-u_{\beta }u_{\mu }g_{\alpha \nu }+u_{\beta }u_{\nu }g_{\alpha \mu })\Big]%
\langle u^{\lambda }{\Theta }_{\lambda \sigma }^{g}u^{\sigma
}\rangle .
\end{eqnarray}
The imaginary part of the spectral density can be extracted by
applying the following equality for $n\geq2$:
\begin{eqnarray}\label{PV}
\Gamma \Big( \frac{D}{2}-n \Big)\Big(-
\frac{1}{L}\Big)^{\frac{D}{2}-n}\rightarrow \frac{(-1)^{n-1}}{
(n-2)!} (-L)^{n-2}\mathrm{ln}(-L)~~
\end{eqnarray}
and then replacing $D\rightarrow 4$ and also, we can adopt the
principal value prescription:
\begin{eqnarray}\label{PV}
\frac{1}{s-m_{X_{b(c)}}}=PV
\frac{1}{s-m_{X_{b(c)}}}-i\pi\delta(s-m_{X_{b(c)}}).
\end{eqnarray}
Then we substitute the propagators into the correlation functions
and related integrals are performed. To remove the contributions
originating from higher states we enforce the standard Borel
transformation in terms of $q^{2}$ in the invariant amplitude,
selecting the structures $g_{\mu \nu }$ and unit matrix for the
axial-vector and scalar states, respectively in both physical and
QCD side, equalizing the attained statement with the related part
of
$\mathcal{\hat{B}}(q^{2})\Pi^{%
\mathrm{Phys}}(q,T)$, finally the pole residue SVZSR for $X_b$ and
$X_c$ particles are extracted at finite temperature:
\begin{eqnarray}\label{ResidueSR}
&&m_{X}^{2}(T)f_{X}^{2}(T)~e^{-m_{X}^{2}(T)/M^{2}}=
\int_{\mathcal{M}^{2}}^{s_{0}(T)}ds\rho^{\mathrm{QCD}}(s,T)~e^{-s/M^{2}}.\quad
\end{eqnarray}
In order to find the mass TSVZSR we should expel the hadronic pole
residue somehow, i.e. taking the derivative of the pole residue
sum rule in Eq.~(\ref{ResidueSR}) in terms of $(-1/M^2)$ and next
dividing by itself, we can reach the thermal mass SVZSR of the
considered hadronic state being $M^2$ the Borel mass parameter and
$s_0(T)$ thermal continuum threshold parameter, respectively:
\begin{equation}\label{massSR}
m_{X}^{2}(T)=\frac{\int_{\mathcal{M}^{2}}^{s_{0}(T)}ds~s~\rho ^{\mathrm{QCD}%
}(s,T)~e^{-s/M^{2}}}{\int_{\mathcal{M}^{2}}^{s_{0}(T)}ds~\rho^{\mathrm{QCD}}
(s,T)~e^{-s/M^{2}}}.
\end{equation}

For compactness, the explicit forms of all spectral densities are
presented in Appendix.\\

\section{Numerical Analysis}\label{sec:NumAnal}
In this section, we find out the numerical values of mass and pole
residue of $X_{b(c)}$ states both at the QCD vacuum and also
$T\neq0$ case. By analyzing the calculations  one can see the hot
medium effects on the hadronic parameters of the investigated
state. During the computations, we used the input parameters in
Table~\ref{tab:input}.

\begin{table}[h!]
\caption{Input parameters
\cite{Tanabashi,Shifman,Reinders,Narison:2010cg}\label{tab:input}}
\centering
\begin{tabular}{ll}
\hline
Parameters &Values\\
\hline
$ m_u=(2.9\pm 0.6)~\mathrm{MeV}$               & $\langle 0| \overline{q}q |0\rangle= -(0.24 \pm 0.01)^3~\mathrm{GeV^3}$   \\
$ m_d=(5.2\pm 0.9)~\mathrm{MeV}$               & $\langle 0|\frac{\alpha_s G^2}{\pi}|0\rangle=(0.022\pm0.004)~\mathrm{GeV}^4 $ \\
$ m_s=(95\pm 5)~\mathrm{MeV}$                  & $\langle 0|\overline{s}s|0 \rangle /\langle 0| \overline{q}q|0\rangle =0.8$ \\
$ m_b=(4.18\pm 0.03)~\mathrm{GeV}$             & $m_{0}^{2}=(0.8\pm0.2)~\mathrm{GeV}^2$ \\
$ m_c=(1.275\pm 0.025)~\mathrm{GeV}$           & \\
\hline
\end{tabular}
\end{table}

In addition to these input parameters, we need the
temperature-dependent quark and gluon condensates, and the energy
density expressions, too. For the thermal quark condensate, the
fit function attained in Ref.~\cite{Dominguez:2016roi} by fitting
Lattice data~\cite{Bali:2012zg} is used, being the light quark
vacuum condensate $\langle 0|\bar{q}q|0\rangle $, thermal version
of quark condensate is determined as:
\begin{eqnarray}\label{eq:qbarqT}
\langle\bar{q}q\rangle_{T}=\langle 0| \bar{q}q |0\rangle(A
e^{\alpha T}+B)^{3/2}.
\end{eqnarray}
In Eq.~(\ref{eq:qbarqT}) $\alpha=0.0412~\mathrm{MeV^{-1}}$,
$A=-6.444\times 10^{-4}$, and $B=0.994$ are coefficients of the
fit function. For the temperature-dependent gluon condensate found
from Lattice QCD data~\cite{Miller:2006hr}, the following
parametrization is employed:
\begin{eqnarray} \label{G2Tfit}
\langle G^{2}\rangle  &=&\langle 0|G^{2}|0\rangle
\Big[C+D\Big(e^{\beta T-\gamma}+1 \Big)^{-1}\Big],
\end{eqnarray}
with the coefficients
$\beta=0.13277~\mathrm{MeV^{-1}}$,~$\gamma=19.3481$, $C=0.55973$
and $D=0.43827$, $\langle 0|G^{2}|0\rangle $ is the gluon
condensate in vacuum state and $G^2 =
G^A_{\alpha\beta}G_A^{\alpha\beta}$. Additionally, the gluonic and
fermionic parts of the energy density parametrization is included
to the calculation achieved in Ref.~\cite{Azizi:2016ddw} from the
Lattice QCD graphics given in Ref.~\cite{Cheng:2007jq}:
\begin{eqnarray}\label{tetamumu}
\langle \Theta _{00}^{g}\rangle &=&\langle \Theta
_{00}^{f}\rangle= \dfrac{1}{2}\langle \Theta_{00}\rangle
\nonumber\\
&=&T^{4}e^{{(\lambda_1 T^{2}-\lambda_2 T)}}+E T^{5}.
\end{eqnarray}
where
$\lambda_1=113.867~\mathrm{GeV^{-2}},\lambda_2=12.190~\mathrm{GeV^{-1}}$
and $E=-10.141~\mathrm{GeV^{-1}}$. To continue the computation one
should also determine the temperature-dependent continuum
threshold for the $X_{b(c)}$ state which is an auxiliary
parameters in the model. The continuum threshold expression is
generated by~\cite{Dominguez:2016roi};
\begin{eqnarray}\label{eq:s0qqbarratio}
\frac{s_0(T)}{s_0}= \bigg[\frac{\langle \bar{q}q
\rangle_T}{\langle 0| \bar{q}q |0\rangle }\bigg]^{2/3},
\end{eqnarray}
where $s_0$ is the continuum threshold at zero temperature.
Actually this parameter is not completely arbitrary but
characterizes the beginning of the first excited state with the
same quantum numbers as the chosen interpolating currents for the
considered particle. The working region for the $s_0$ is
determined so that the physical quantities show relatively weak
dependence on it.

\begin{figure}[h!]
\begin{center}
\includegraphics[width=0.8\textwidth]
{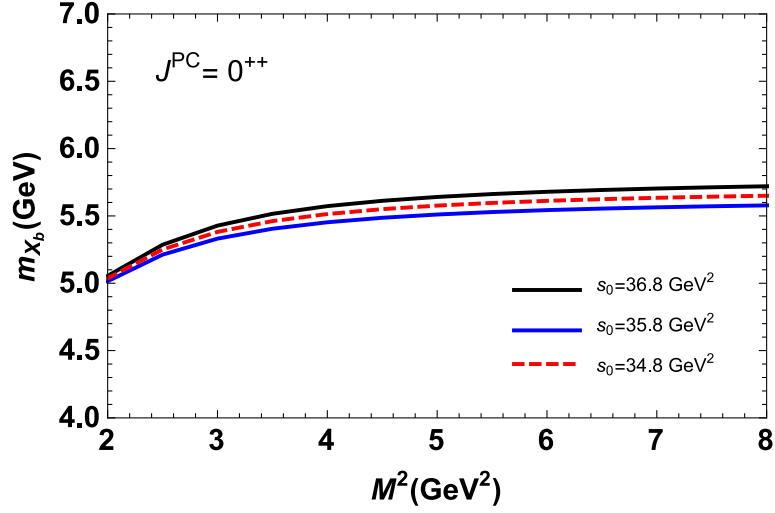} \caption{The mass of $X_b$ state versus the
Borel mass parameter $M^2$.}\label{fig:MassMsq}
\end{center}
\end{figure}

Next, we discuss the employed parameter region of continuum
threshold $s_0$ and Borel mass parameter $M^2$, which is mainly
restricted by the convergence of the OPE. The idea of the SVZSR
method dictate us that the physical quantities should be
independent of the continuum threshold $s_0$ and Borel mass
parameter $M^2$.  After some analyzes, we defined the range of
Borel parameter $M^2$ and continuum threshold $s_0$ such that
hadronic parameters are stable at these intervals. We looked for
the OPE convergence and the pole contribution dominance and
determined the conventional Borel window in the SVZSR approach to
ensure the quality of the analysis. The lower bound of the Borel
parameter $M_{\mathrm{min}}^{2}$ is fixed from convergence of the
OPE. By quantifying this constraint we require that contributions
of the last terms, that is dimension five plus six, in OPE are
found around $15\%$.
\begin{equation}\label{eq:Dim6Contribution}
\frac{\Pi ^{(\mathrm{Dim5}+\mathrm{Dim6})}(M_{\mathrm{min}}^{2},\ s_{0})}{\Pi (M_{\mathrm{%
min}}^{2},\ s_{0})}\cong 0.15.
\end{equation}
We also get an upper limit constraint for $M_{\mathrm{max}}^{2}$
by imposing the severe constraint that the QCD continuum
contribution must be smaller than the pole contribution:
\begin{equation}\label{eq:PC}
\mathrm{PC}(s_0, M^2)=\frac{\Pi (M_{\mathrm{max}}^{2},\ s_{0})}{\Pi (M_{\mathrm{max}%
}^{2},\ \infty )}>\frac{1}{2}.
\end{equation}
Finally, a working region for $M^2$ and $s_0$ are fixed according
to the above mentioned criteria, thus we arrive the following
interval for the $X_b$:
\begin{eqnarray*}
~~M^2 \in [4-6]~\mathrm{GeV}^2~~;~~~~s_0 \in
[34.8-36.8]~\mathrm{GeV}^2,
\end{eqnarray*}
and for the $X_c$ state:
\begin{eqnarray*}
M^2 \in [2-4]~\mathrm{GeV}^2~~;~~~~s_0 \in
[8.6-9.8]~\mathrm{GeV}^2.
\end{eqnarray*}
In this region the dependence of the mass and pole residue on
$s_0$ and $M^2$ is fixed anymore, and we guarantee that the sum
rules give the reliable results. Plotting the mass versus $M^2$ at
different fixed values of the continuum threshold $s_0$ in figure
\ref{fig:MassMsq}  at $T=0$ we see the independence of mass from
$M^2$.
Numerical results obtained for the mass and pole residue in vacuum
are shown in Table~\ref{tab:Xbscalar}-Table~\ref{tab:Xcaxial} and
our results are consistent with the results existing in the
literature~\cite{D0:2016mwd,Chen:2016mqt,Agaev:2016mjb,Agaev:2016lkl}.\\

%
\begin{table}[h!]
\begin{center}
\caption{Comparison of the mass and pole residue vacuum values of
$X_{b}$ for the ``scalar case'' with theoretical models and
experimental results available in the
literature.}\label{tab:Xbscalar}
\begin{tabular}{lcccc}
\hline
Parameter                           & $m_{X_b}(\mathrm{MeV})$                         & $f_{X_b}(\mathrm{GeV^4})$   \\
\hline
Present Work                        & $5567^{+112}_{-114}$                            & $(0.35^{+0.07}_{-0.06})\times 10^{-2}$  \\
Experiment                          & $5567.8 \pm 2.9$~\cite{D0:2016mwd}              & $ -$    \\
RQM                                 & $5997~\cite{Ebert:2010af}$                      & $ -$    \\
NRQM                                & $~~~~5980~\cite{Ghalenovi:2015mha}~\mathrm{or}$ & $ -$    \\
                                    & $5901~~~~~$                                     & $ -$    \\
SVZSR                               & $5580 \pm 140$~\cite{Chen:2016mqt}              & $ -$      \\
SVZSR                               & $5584 \pm 137$~\cite{Agaev:2016mjb}             & ~~~~~~~~~$(0.24 \pm 0.02)\times10^{-2}$~\cite{Agaev:2016mjb}\\
\hline
\end{tabular}
\end{center}
\end{table}
\begin{table}[htbp]
\begin{center}
\caption{Comparison of the mass and pole residue vacuum values of
$X_{c}$ for the ``scalar case'' with theoretical models and
experimental results available in the
literature.}\label{tab:Xcscalar}
\begin{tabular}{lcccc}
\hline
Parameter                     & $m_{X_c}(\mathrm{MeV})$                   & $f_{X_c}(\mathrm{GeV^4})$       \\
\hline
Present Work                  & $2675^{+128}_{-131}$                      & $(0.39^{+0.07}_{-0.06})\times 10^{-2}$  \\
Experiment                    & $ - $                                     & $ - $    \\
RQM                           & $2619~\cite{Ebert:2010af}$                & $ - $    \\
SVZSR                         & $2550 \pm 90$~\cite{Chen:2016mqt}         & $ - $   \\
SVZSR                         & $2634 \pm 62$~\cite{Agaev:2016lkl}        & ~~~~$(0.11 \pm 0.02)\times 10^{-2}$    \\
\hline
\end{tabular}
\end{center}
\end{table}

\begin{figure}[h!]
\begin{center}
\includegraphics[width=0.49\textwidth]{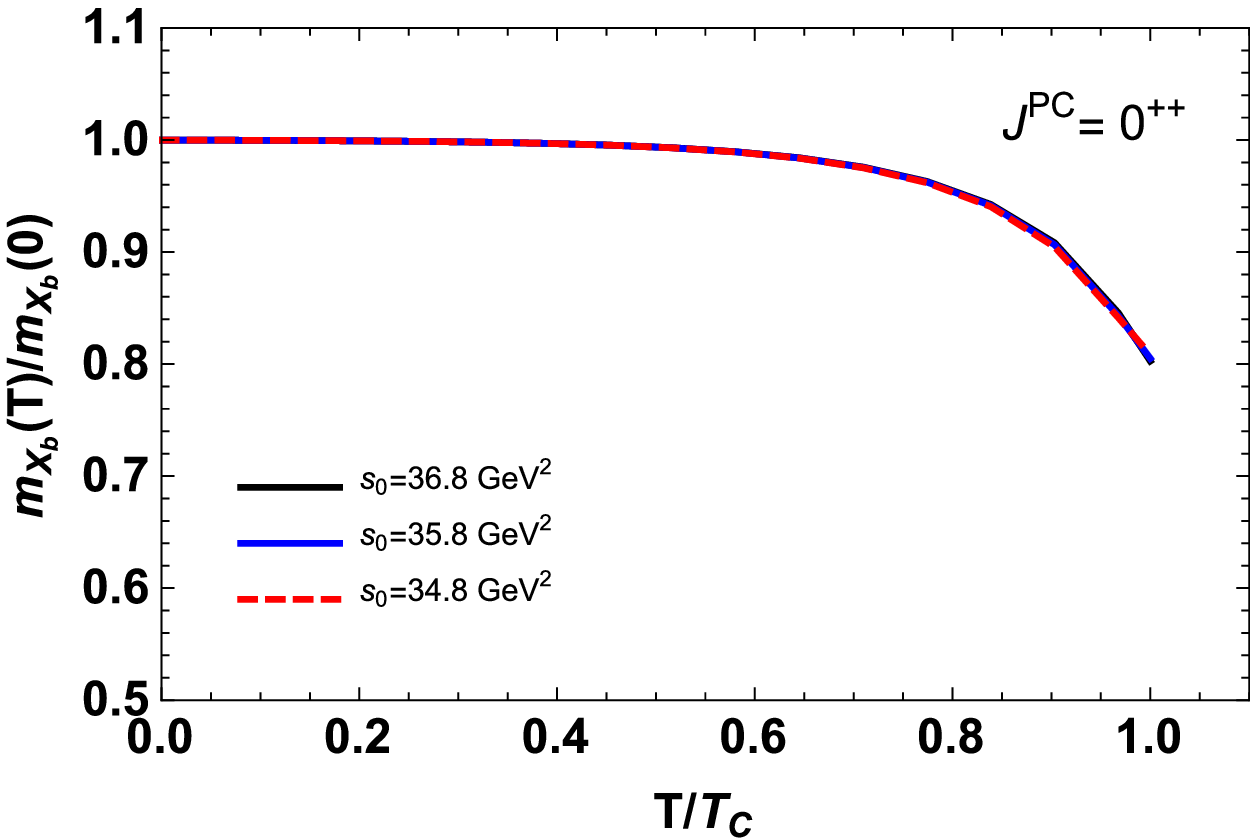}\hskip0.1cm\includegraphics[width=0.49\textwidth]
{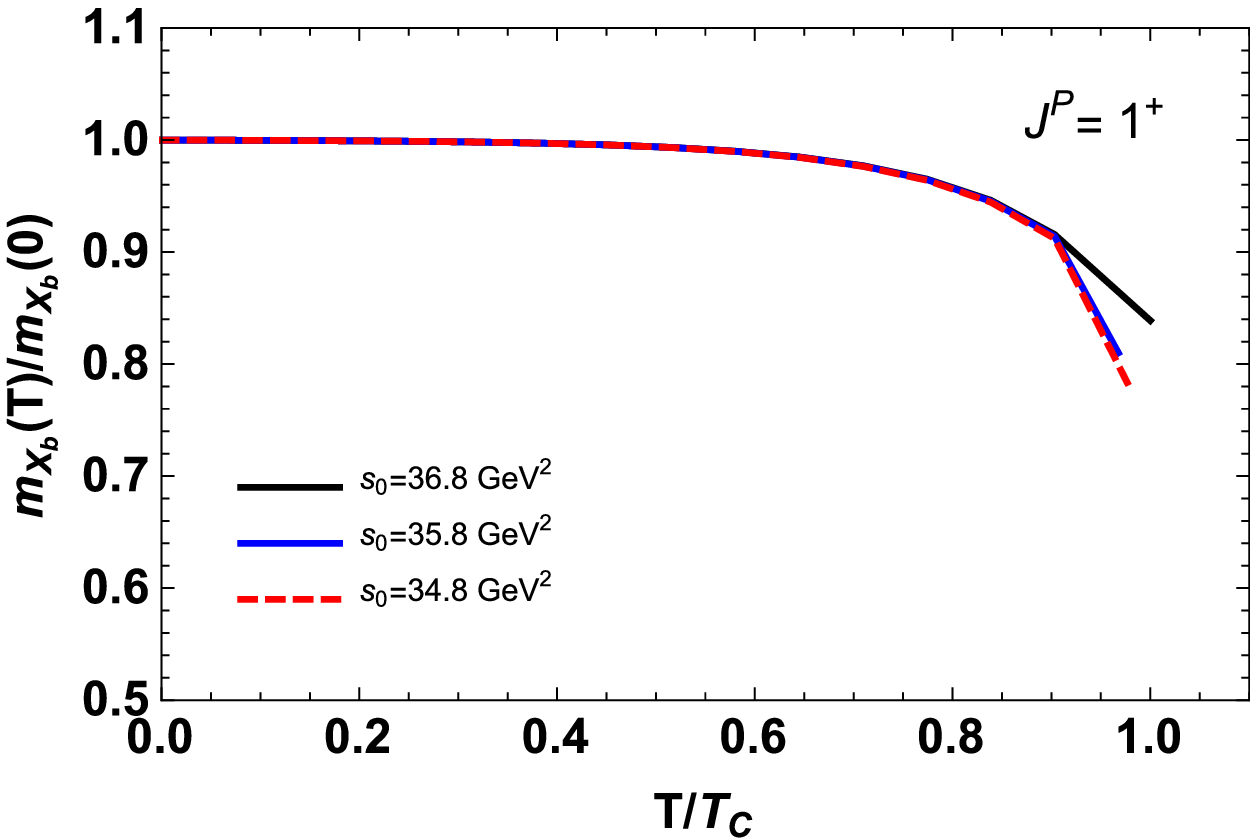} \caption{Mass changes as a function of
 temperature of scalar (Left) and axial-vector
$X_b$ state (Right).}\label{fig:MassVsTempXb}
\end{center}
\end{figure}

\begin{figure}[h!]
\begin{center}
\includegraphics[width=0.49\textwidth]{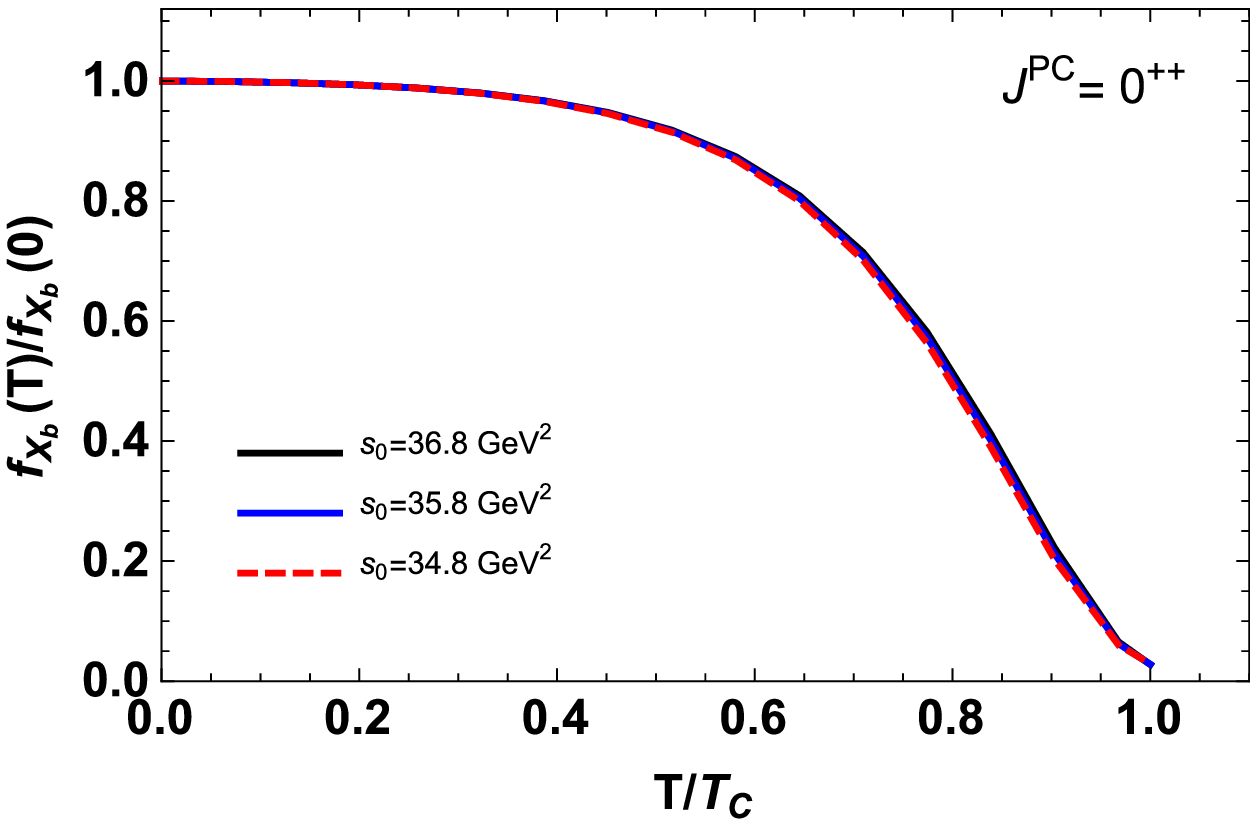}\hskip0.1cm\includegraphics[width=0.49\textwidth]
{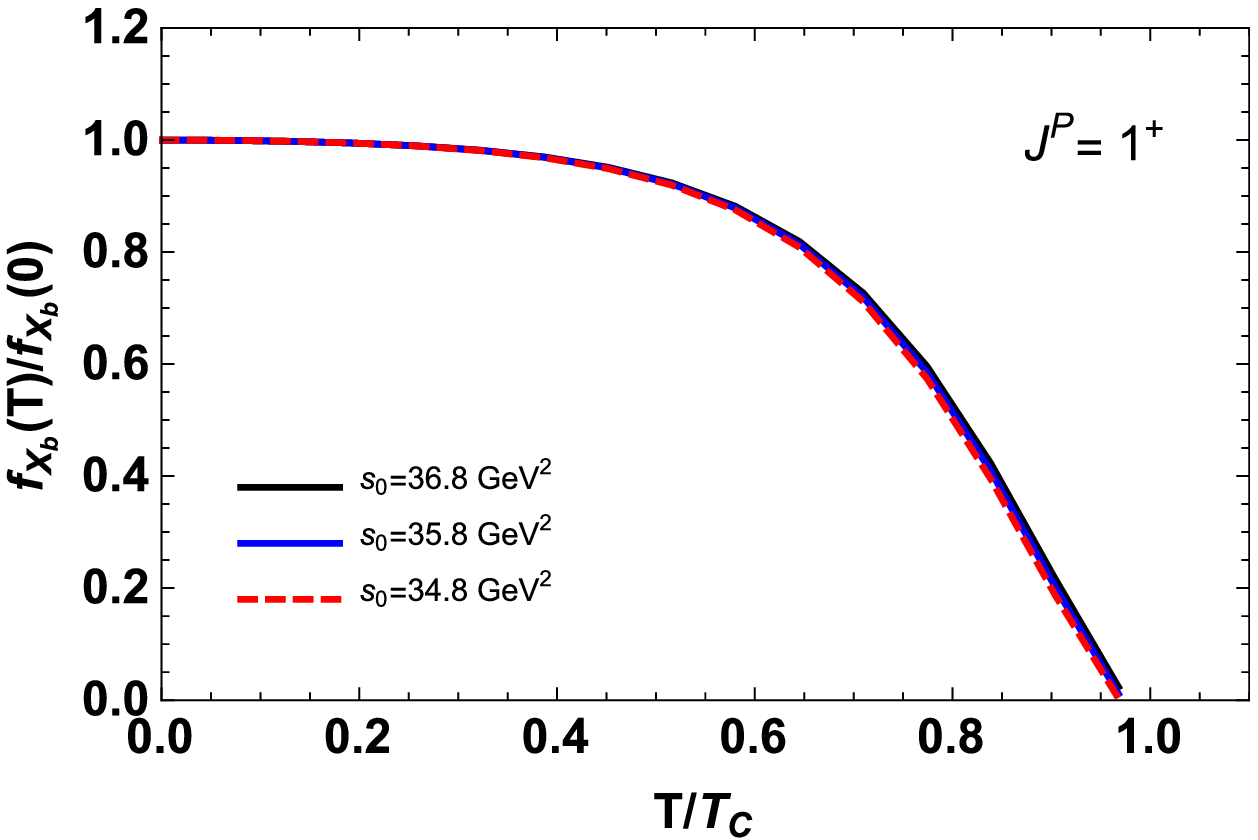} \caption{Pole residue variations as a function
of temperature of scalar (Left) and axial-vector $X_b$ state
(Right).}\label{fig:ResidueVsTempXb}
\end{center}
\end{figure}

\begin{figure}[h!]
\begin{center}
\includegraphics[width=0.49\textwidth]{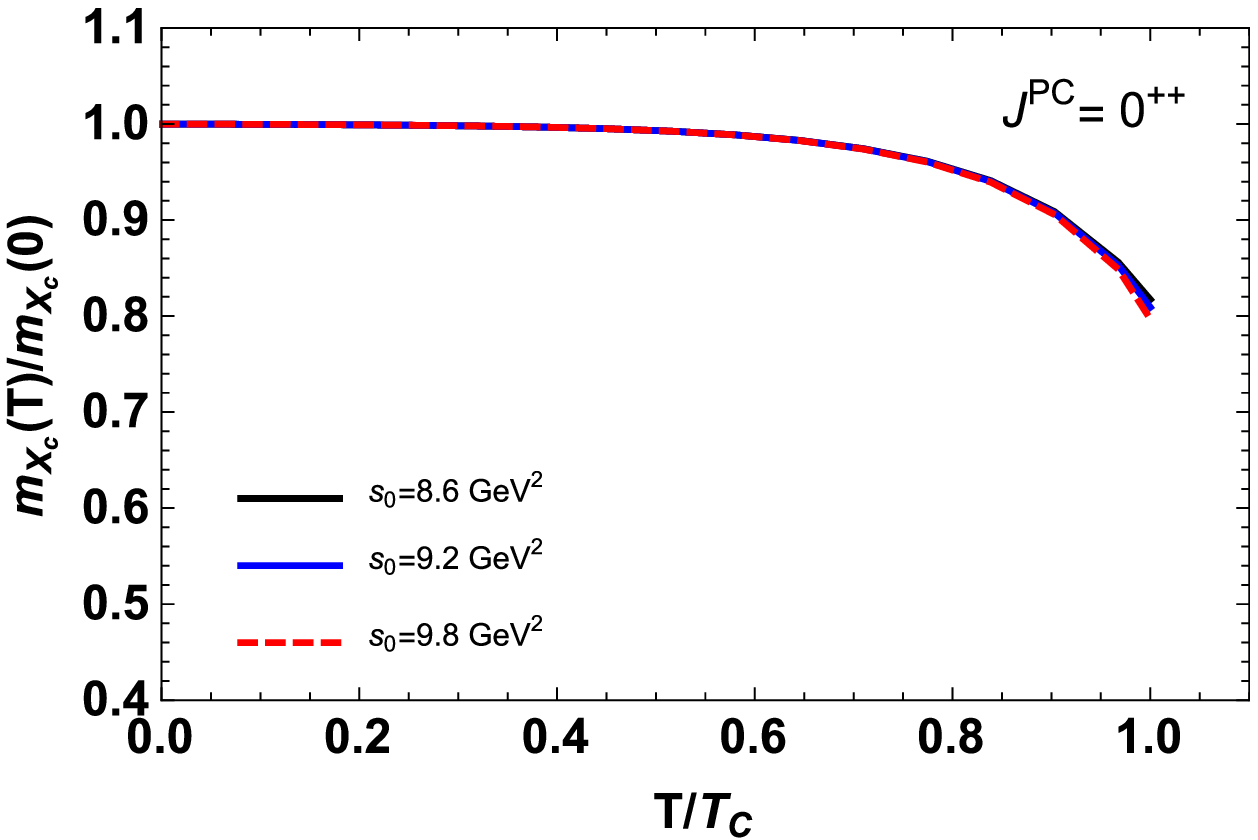}\hskip0.1cm\includegraphics[width=0.49\textwidth]
{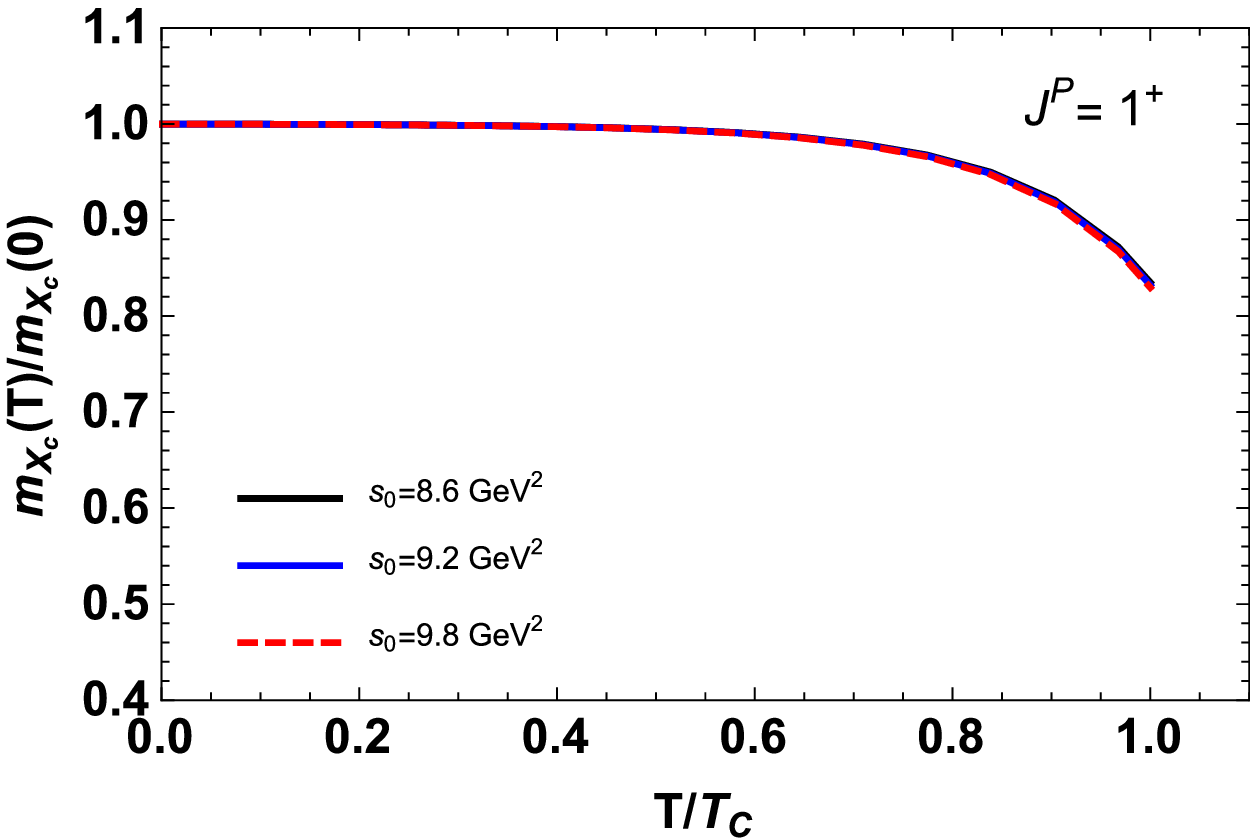} \caption{ Mass changes in terms of temperature
of scalar (Left) and axial-vector $X_c$ state
(Right).}\label{fig:MassVsTempXc}
\end{center}
\end{figure}

\begin{figure}[h!]
\begin{center}
\includegraphics[width=0.49\textwidth]{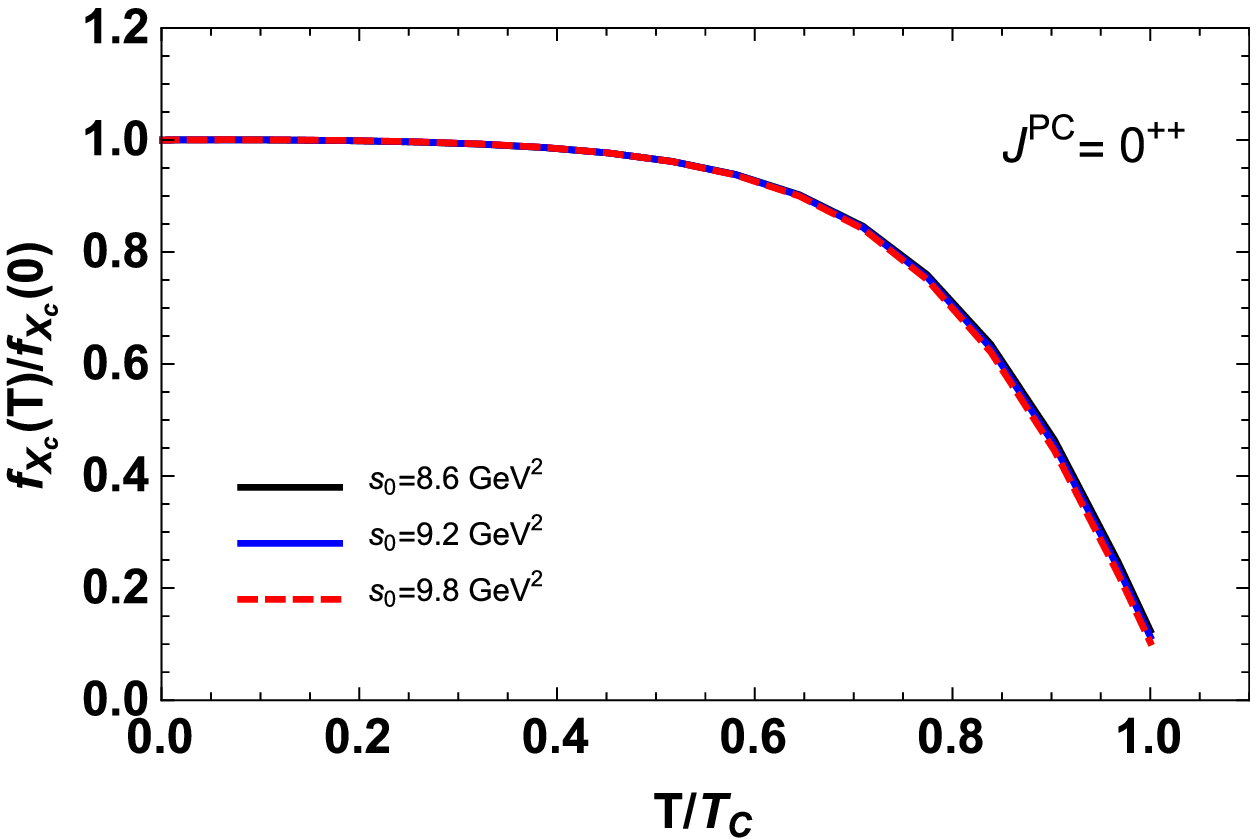}\hskip0.1cm\includegraphics[width=0.49\textwidth]
{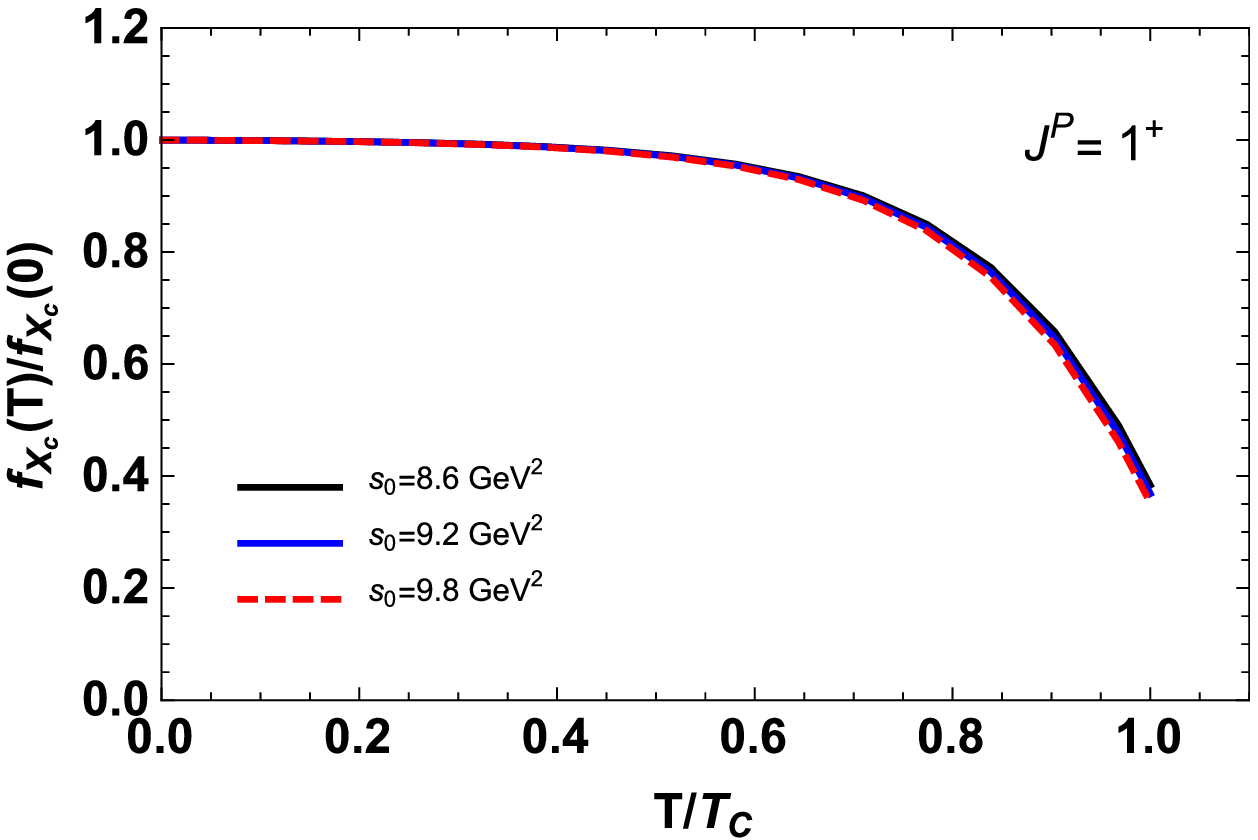} \caption{Pole residue variations in terms of
temperature of scalar (Left) and axial-vector $X_c$ state
(Right).}\label{fig:ResidueVsTempXc}
\end{center}
\end{figure}

\begin{table}[htbp]
\begin{center}
\caption{Comparison of the mass and pole residue vacuum values of
$X_{b}$ for the ``axial case'' with theoretical models and
experimental results available in the
literature.}\label{tab:Xbaxial}
\begin{tabular}{lcccc}
\hline
Parameter                       & $m_{X_b}(\mathrm{MeV})$                   & $f_{X_b}(\mathrm{GeV^4})$   \\
\hline
Present Work                    & $5569^{+103}_{-102}$                      & $(0.22^{+0.04}_{-0.03})\times 10^{-2}$   \\
Experiment                      & $5567.8 \pm 2.9$~\cite{D0:2016mwd}        & $ -$    \\
RQM                             & $6125~\cite{Ebert:2010af}~\mathrm{or}$    & $ - $    \\
                                & $6021~~~~~~~~~$                           & $ - $    \\
SVZSR                           & $5590 \pm 150$~\cite{Chen:2016mqt}        & $ -$      \\
SVZSR                           & $5864 \pm 158$~\cite{Agaev:2016mjb}       & ~~~~~~~~$(0.42 \pm 0.14)\times10^{-2}$~\cite{Agaev:2016mjb}\\
\hline
\end{tabular}
\end{center}
\end{table}
\begin{table}[h!]
\begin{center}
\caption{Comparison of the mass and pole residue vacuum values of
$X_{c}$ for the ``axial case'' with theoretical models and
experimental results available in the
literature.}\label{tab:Xcaxial}
\begin{tabular}{lcccc}
\hline
Parameter                     & $m_{X_c}(\mathrm{MeV})$                   & $f_{X_c}(\mathrm{GeV^4})$       \\
\hline
Present Work                  & $2557^{+124}_{-122}$                      & $(6.08^{+0.78}_{-0.74})\times 10^{-2}$  \\
Experiment                    & $ - $                                     & $ - $    \\
SVZSR                         & $2550 \pm 100$~\cite{Chen:2016mqt}        & $ - $   \\
\hline
\end{tabular}
\end{center}
\end{table}
Our last target is to look for the variations of the mass and pole
residue of the $X_b$ and $X_c$ resonances in terms of temperature.
Mass and pole residue versus temperature plots are drawn in
figures~\ref{fig:MassVsTempXb}-\ref{fig:ResidueVsTempXc}.

This graphs display that the mass and pole residue of the $X_b$
state stay roughly unmodified until $T \cong
0.12~\mathrm{\mathrm{GeV}}$, nonetheless, after this point, they
begin to decrease promptly with increasing temperature. However
diminishing of the mass and pole residue value with the
temperature does not mean a stability of the studied state. To
make a general deduction on the stability of the particle one
should compute its decay width as well. Actually, similar to the
mass and pole residue, the decay width of the particle depends
also on the temperature. For instance in
Ref.~\cite{Dominguez:2009mk,Azizi:2010zza} despite decreasing of
the considered particles' mass and pole residue in terms of
temperature, decay widths are increased with the temperature.
%
\section{Conclusion}\label{sec:Result}

In this work, we have revisited the bottomonium and charmonium
states $X_b$ and $X_c$ extending our model from vacuum state to
heat bath. To describe the effects of hot medium to the hadronic
parameters of the resonances $X_b$ and $X_c$, Thermal SVZSR model
is used considering contributions of condensates up to dimension
six. We hope that renew interpretation of $X_b$ resonance in hot
medium may give different insights for understanding the inner
structure of unfitted bottomonium states with Quark Model. Due to
its observed decay mode, the  $X_b$ must contain four different
valence quark components, which makes the $X_b$ a good candidate
for a tetraquark state.

We investigate this state in axial-vector and scalar picture as
tetraquark candidate. Numerical findings show that the  $X_b$ can
be well described by both scalar and axial-vector tetraquark
currents. This particle has almost equal possibility for being a
scalar or axial-vector particle. Our results at $T=0$ are in
reasonable agreement with the available experimental data and
other SVZSR works in the literature. The exact result can only be
determined by the precise measurement of the decay width values by
the experiments. Additionally, our numerical calculations indicate
that the mass and pole residue values of the considered states are
stable at low temperatures, but they reduce by roughly $20\%$ and
$98\%$ of their vacuum values for the  $X_b$ state and also for
the charmed partner $20\%$ and $90\%$, respectively when the
temperature approaches to phase transition temperature for the
scalar assumption. In the axial-vector picture, these values
decrease by $17\%$ and $99\%$ of their vacuum values for the
$X_b$, $18\%$ and $65\%$ of its charmed partner, too.

There are some comments on that this decrease can indicate the
deconfinement phase transition in quark-gluon plasma which also
occur in the early universe. In the literature, remarkable drop in
the values of mass and pole residue in hot medium can be regarded
as the signal of the quark-gluon plasma (QGP), called as new state
of matter, phase transition. Also, the manner of  $X_b$ state
according to temperature can be a useful tool to analyze the
heavy-ion collision experiments. Our estimates for the hadronic
features of the  $X_b$ meson can be tested in the forthcoming
experiments such as CMS, LHCb and PANDA.

The  $X_b$ data will provide a rich physics output and this makes
a motivated issue for the Belle-II initial data taking. We hope
that precise spectroscopic measurements are predicted at the
Super-B factories and at the LHC will supply conclusive answers to
open questions raised here such as unconventional quark
combinations, interactions in exotic hadrons, etc. and will help
resolve the current and long-standing puzzles in the exotic
bottomonium and charmonium sectors.

However, the production mechanism of the  $X_b$ is very different
at the $p\overline{p}$ and $pp$ colliders. Future experimental
efforts are desirable in the clarification of the situation on the
 $X_b$ state and its charmed partner. In 2004 SELEX
Collaboration~\cite{Cooper:2004hj} reported the first observation
of a charm-strange meson $D^{+}_{sJ}(2632)$ at a mass of
$2632.5\pm 1.7 ~\mathrm{MeV/c^2}$, the charm hadro-production
experiment E781 at Fermilab. Since this particle has nearly the
same mass as the $X_c$ that was discovered in the SELEX
experiment, it may most likely be the same particle.

Further detailed experimental and theoretical studies of the
invariant mass of $B^0_s\pi^\pm$ spectrum, production and decays
of tetraquark states with four different flavors in the future are
severely called for towards a better understanding their nature
and the classification of exotics.

\appendix

\section{The spectral densities} \label{sec:App}

\renewcommand{\theequation}{\Alph{section}.\arabic{equation}} \label{sec:App}

In this appendix the results of the spectral densities in our
calculations are presented. The spectral density can be written
separating the terms according to the operator dimensions as:
\begin{eqnarray}\label{Rhoall}
\rho^{\mathrm{QCD}}(s,T)=\rho^{\mathrm{pert.}}(s)+\rho^{\mathrm{non-pert.}}(s,T).
\end{eqnarray}
Here,
\begin{eqnarray}\label{Rhononpert}
\rho^{\mathrm{non-pert}}(s,T)&=&\rho^{\langle\bar{q}q\rangle}(s,T)+\rho^{\langle
G^2\rangle+\langle \Theta_{00}\rangle}(s,T)\notag \\
&+&\rho^{\langle\bar{q}Gq\rangle}(s,T)+\rho^{\langle\bar{q}q\rangle^2}(s,T).
\end{eqnarray}
The complete expressions for $\rho^{\mathrm{pert.}}(s)$ and
$\rho^{\mathrm{nonpert.}}(s,T)$ are shown below as the integrals
over the Feynman parameter $z$ for the axial assumption
($J^P=1^+$) of $X(5568)$:
\begin{eqnarray}\label{RhoPertaxial}
\rho^{\mathrm{pert.}}(s)&=& \frac{1}{3 \times 2^{12}~
\pi^6}\int_{0}^{1}dz \frac{1}{r^3}\Bigg\{ z^2 (r s + m_{b}^2)^2
\Bigg[z \Big(5 s^2 z \zeta + 6 \phi s m_{b}^2 + z m_{b}^4 \notag \\
&-& 16 m_b m_d (r s + m_{b}^2) \Big) +16 m_s m_u \Big(4 s z \zeta
+\phi m_{b}^2 - 9 r m_b m_d \Big) \Bigg] \Bigg\}\notag
\\&\times& \theta[L(s,z)],
\end{eqnarray}
\begin{eqnarray}\label{RhoDim3axial}
&&\rho^{\langle q\bar{q}\rangle}(s,T)= \frac{1}{2^7~
\pi^4}\int_{0}^{1}dz \frac{1}{r^2}\Bigg\{ z \Bigg[r (r s + m_b^2)
\big(3 \phi s + z m_b^2 - 4 m_b m_d \big)\notag \\
&&\times \bigg\{ m_s \Big(\langle s\bar{s} \rangle- 2\langle
u\bar{u}\rangle\Big)+ m_u \big(-2 \langle s\bar{s} \rangle+\langle
u\bar{u} \rangle \big)  \bigg\} +\langle d\bar{d} \rangle \Big[2 z
m_b^5+\phi m_b^4 m_d \notag \\
&&+ 4 \zeta m_b^2 m_d (s z + m_s m_u) + 4 r m_b^3 (s z + 2 m_s
m_u)+ 2 s \zeta m_b (s z + 4 m_s m_u) \notag \\&&+ s \varphi m_d
(3 s z+8m_sm_u)\Big]\Bigg]\Bigg\}~\theta[L(s,z)],
\end{eqnarray}
\begin{eqnarray}\label{RhoDim4axial}
&&\rho^{\langle G^2 \rangle+\langle\Theta_{00}\rangle}(s,T)=
\frac{1}{\pi^4}\Big\langle\frac{ \alpha_s G^2}{\pi}\Big\rangle
\int_{0}^{1}dz  \frac{1}{9\times 2^5 } \Bigg\{\frac{1}{r^3} z
\Bigg[4 s^2 \varphi z (-27 + 32 z) \notag \\
&&+ m_b \Bigg\{2\phi s m_b
\bigg(72 + z (-160 + 89z)\bigg) + (6 - 7 z)^2 m_b^3 - 4 \zeta m_d \Big[s \big(-36\notag \\
&& + z (36 +
z)\big) +36 m_b^2\Big]\Bigg\} + 4\phi m_s m_u \Big(24 s \varphi  + (-18 + 19 z) m_b^2 \Big) \Bigg]~\theta[L(s,z)] \notag \\
&&+ \frac{\langle \Theta^f_{00} \rangle}{32 r} \Bigg[ z \bigg(s^2
z 3 + 80 z\bigg) \zeta +  m_b \bigg\{72 \phi s z m_b + z (-3 + 8
z)
m_b^3 - 24 r  m_d \notag \\
&&\bigg(s (-1 + 3 z) + m_b^2 \bigg)  \bigg\}+ 12 r m_s m_u(4 r s +
m_b^2) \Bigg]\notag \\
&&+\frac{\langle \Theta^g_{00} \rangle}{8 \pi^2 r^2} g_s^2 z \zeta
\Big[s^2 z (-1 +
4 z) (-9 + 10 z)  + m_b \bigg\{4 \phi s m_b \bigg(3 + z (-11 + 9 z) \bigg)\notag \\
&&+ z (3 - 6 z + 4 z^2) m_b^3 - 12 \zeta \bigg[s m_d \bigg((-1 + 3
z) + m_b^2 \bigg)\bigg]  \bigg\} \notag \\
&&+ 6 \phi  m_s m_u
\big(2 r s + m_b^2\big) \Big]\Bigg\},
\end{eqnarray}
\begin{eqnarray}\label{RhoDim5part1and2axial}
&&\rho^{\langle \bar{q}G q\rangle}(s,T)= \frac{1}{3\times 2^6
\pi^4}\int_{0}^{1}dz \frac{m_0^2}{r} \Bigg\{ r \bigg(2 \phi s + z
m_b^2 - m_b m_d\bigg) \Big[m_s \big(\langle s\bar{s} \rangle-
3 \notag \\
&&\times\langle u\bar{u} \rangle\big) + m_u \big(-3 \langle
s\bar{s} \rangle+\langle u\bar{u} \rangle\big)  \Big]+ \langle
d\bar{d} \rangle \bigg[3 z m_b^3 + \phi m_b^2 m_d + 3 r m_b (s z +
m_s \notag \\
&& \times m_u) + \zeta m_d (2 s z + m_s m_u)\bigg]\Bigg\}
\end{eqnarray}
and
\begin{eqnarray}\label{RhoDim6part1and2axial}
&&\rho^{\langle \overline{q}q \rangle^2}(s,T)= \frac{1}{3\times
2^4 \pi^2}\int_{0}^{1}dz \Bigg\{\frac{1}{27\pi^2} \Big[g^2_{s}
\bigg(\langle s\bar{s}\rangle^2 +\langle u\bar{u} \rangle^2 \bigg)
(2 \phi s + z m_b^2 \notag \\
&&- m_b m_d) +27\langle d\bar{d} \rangle \pi^2 (2 m_b + r m_d)
\bigg[ m_s\big(\langle s\bar{s} \rangle- 2\langle u\bar{u} \rangle
\big)  + m_u \big(-2 \langle s\bar{s} \rangle \notag \\
&& +\langle u\bar{u} \rangle \big)  \bigg]+ \langle d\bar{d}
\rangle^2 g^2_{s}(2 r s z + z m_b^2 + r m_s m_u) + 27 \pi^2
\langle s\bar{s}\rangle \langle u\bar{u}\rangle \big(8 r s z+ 4 z \notag \\
&&\times  m_b^2 - 4 m_b m_d + r m_s m_u
\big)\Big]\Bigg\}~\theta[L(s,z)]+ \langle s\bar{s} \rangle \langle
u\bar{u}  \rangle m_b m_d m_s m_u \delta(s-m_b^2).\notag \\
\end{eqnarray}
For the scalar assumption ($J^{PC}=0^{++}$) we get the following
expressions for the spectral density as follows:
\begin{eqnarray}\label{RhoPertscalar}
&&\rho^{\mathrm{pert.}}(s)= \frac{1}{3\times 2^9 \pi^6}
\int_{0}^{1}dz \frac{1}{r^3} \Bigg\{\bigg[z^2 (r s + m_b^2)^2
\bigg\{-z \bigg(3 s^2 z \zeta + 4 \phi s m_b^2 + z m_b^4 \bigg) +\notag \\
&& 4z m_b m_d (r s+ m_b^2) - 4 m_s m_u \bigg(5 s z \zeta + 2 m_b
\big(\phi m_b - 9 r m_d \big)\bigg)
\bigg\}\Bigg]\Bigg\}~\theta[L(s,z)],
\end{eqnarray}
\begin{eqnarray}\label{RhoDim3scalar}
&&\rho^{\langle q\bar{q}\rangle}(s,T)=-\frac{1}{2^5 \pi^4}
\int_{0}^{1}dz \frac{1}{r^2}\Bigg\{ z \Bigg[2 r (r s + m_b^2)
\Bigg\{2 \phi s \bigg(\langle s\bar{s} \rangle- \langle u\bar{u}
\rangle \bigg) (m_s - m_u) \notag \\
&&+\bigg(\langle s\bar{s} \rangle- \langle u\bar{u} \rangle \bigg)
z m_b^2 (m_s - m_u)- m_b m_d \bigg(m_s\big(\langle s\bar{s}\rangle
- 4\langle u\bar{u} \rangle \big)  + m_u\big(-4 \langle s\bar{s}
\rangle\notag \\
&&+\langle u\bar{u} \rangle\big)\bigg)\Bigg\}+ \langle d\bar{d}
\rangle \bigg[z m_b^5 + 2 \phi m_b^4 m_d +2 \zeta m_b^2 m_d (3 s z
+ 2 m_s m_u) + 2 s \varphi m_d (2 s z \notag \\
&& + 3 m_s m_u) + 2 r m_b^3 (s z + 4 m_s m_u) + s m_b \zeta (s z +
8 m_s m_u)\bigg]\Bigg]\Bigg\}~\theta[L(s,z)],
\end{eqnarray}
\begin{eqnarray}\label{RhoDim4scalar}
&&\rho^{\langle G^2 \rangle+\langle \Theta_{00}\rangle}(s,T)=
\frac{1}{6^2\pi^4}\big\langle\frac{ \alpha_s G^2}{\pi}\rangle \big
.\int_{0}^{1} dz \frac{1}{64} \Bigg\{\frac{1}{r^3} \Big[-2 z m_b^4
\bigg(18 + z (-30 + 13 z)\bigg)\notag \\
&& + 2 \zeta s m_b m_d \bigg(36 + (-54 + z) z \bigg)  +36 m_b^3
m_d r (2 - 3 z)  - 12 s \varphi z (-6 s + 4 s z  \notag \\
&&+ 9 m_s m_u)+ \phi m_b^2 \bigg(-3 s(6 - 5 z)^2 + 4 m_s m_u(18 -
19 z) \bigg)\Big]+\frac{\langle  \Theta^f_{00} \rangle}{r^2}\Bigg[
z \bigg(r s^2 \notag \\
&&\times(3 - 50 z) z \zeta + 3 m_b \bigg\{\phi r s  m_b(3 - 16 z)
- 2 z \zeta
m_b^3 +2 m_d \bigg(r^2 s (-1 + 3 z) + \zeta m_b^2 \bigg) \bigg\}\notag \\
&& - 9 \zeta r s  m_s m_u \bigg)\Bigg]-\frac{\langle \Theta^g_{00}
\rangle}{14 \pi^2 r^2}\Bigg[ g_s^2 z \bigg( \phi s m_b^2\bigg(3 +
4 z (-3 + 2 z)\bigg)  + 3 z \zeta m_b^4 +3 r s m_b m_d \notag \\
&&\times \bigg(2 + z (-9 + 8 z)\bigg) + m_b^3 m_d \bigg(6
+ z (-15 + 8 z)\bigg)  + s z \zeta  \bigg\{s [6 + z (-34 \notag \\
&&+ 25 z) + 9 m_s m_u
\bigg\}\bigg)\Bigg]\Bigg\}~\theta[L(s,z)]-\frac{1}{r}\Big\langle\frac{
\alpha_s G^2}{\pi}\Big\rangle m_b m_d m_s m_u s z^2 \delta \big
(s+\frac{m_b^2}{r}\big),
\end{eqnarray}
\begin{eqnarray}\label{RhoDim5scalarpart1and2}
&&\rho^{\langle \bar{q}G q\rangle}(s,T)= \frac{1}{3 \times 2^5
\pi^4} \int_{0}^{1}dz \frac{m_{0}^2}{r} \Bigg\{3\langle d\bar{d}
\rangle z m_b^3 + 3 \phi r s \bigg[ m_s\bigg(2 \langle s\bar{s}
\rangle - 3\langle u\bar{u} \rangle \bigg)+ m_u \notag \\
&&\times\bigg(-3 \langle s\bar{s} \rangle+ 2\langle u\bar{u}
\rangle \bigg) \bigg]+ 2\langle d\bar{d} \rangle \zeta m_d (3 s z
+ m_s m_u) + m_b^2 \Bigg[4\langle d\bar{d} \rangle \phi m_d + 2 r
z  \bigg\{ m_s\notag \\
&&\times \big(2 \langle s\bar{s} \rangle- 3\langle u\bar{u}
\rangle \big) + m_u \big(-3 \langle s\bar{s} \rangle+ 2\langle
u\bar{u}\rangle\big) \bigg\}\Bigg]+ r m_b \Bigg[-m_d
\Big\{m_s\bigg(\langle s\bar{s} \rangle- 6\langle u\bar{u}
\rangle\bigg) \notag \\
&& + m_u \big(-6 \langle s\bar{s} \rangle+\langle u\bar{u}
\rangle\big) \Big\} + 3\langle d\bar{d} \rangle \bigg(s z + 2 m_s
m_u \bigg)\Bigg]\Bigg\}~\theta[L(s,z)],
\end{eqnarray}
and
\begin{eqnarray}\label{RhoDim6scalarpart1and2}
&&\rho^{\langle \overline{q}q \rangle^2}(s,T)= -\frac{1}{3\times
2^3\pi^2} \int_{0}^{1}dz  \frac{1}{27 \pi^2}  \Bigg\{g_{s}^2
\bigg(\langle s\bar{s}\rangle^2 + \langle u\bar{u} \rangle^2\bigg)
(6 \phi s + 4 z m_b^2 - m_b m_d) \notag \\
&& + 2\langle d\bar{d} \rangle^2 g_s^2 \bigg(3 r s z + 2 z m_b^2+
r m_s m_u \bigg) + 108 \pi^2 \langle s\bar{s} \rangle \langle
u\bar{u}\rangle \big(3 r s z + 2 z m_b^2 - 2 m_b m_d \notag \\ &&+
r m_s m_u\big) + 54\langle d\bar{d}\rangle \pi^2 \bigg[ 2 r m_d
(m_s - m_u)\bigg(\langle s\bar{s} \rangle-\langle u\bar{u} \rangle
\bigg) +m_b \bigg\{m_s \bigg(\langle s\bar{s} \rangle- 4\langle
u\bar{u} \rangle \bigg) \notag \\ && +  m_u \bigg(-4 \langle
s\bar{s} \rangle+ \langle u\bar{u} \rangle  \bigg) \bigg\}\bigg]
\Bigg\}~\theta[L(s,z)]-\langle s\bar{s} \rangle \langle u\bar{u}
\rangle m_b m_d m_s m_u \delta(s - m_{b}^2)
\end{eqnarray}
where the explicit expression of the function $L(s,z)$ is
\begin{equation}
L(s,z)=s z (1-z)-z m_{b}^2.
\end{equation}
In the expressions above the following abbreviations is used for
simplicity:
\begin{eqnarray}
\varphi &=&(z-1)^3,  \nonumber \\
\zeta &=& (z-1)^2,  \nonumber \\
\phi &=&z(z-1),  \nonumber \\
r &=& z-1.
\end{eqnarray}
If one makes the replacement $m_b \rightarrow m_c$, the spectral
density of the charmed partner of $X(5568)$ can be easily
obtained.

\end{document}